\documentclass[aps,prb,tightenlines,showpacs,twocolumn,superscriptaddress]{revtex4-1}  %

\usepackage{amsmath,amssymb,amsfonts,bm}
\usepackage{graphicx}
\usepackage{dcolumn}
\usepackage{color}
\usepackage[colorlinks=true,linkcolor=blue,citecolor=blue]{hyperref}%

\begin{document}


\title{Thermodynamics of Two-impurity Anderson Model with Dzyaloshinskii-Moriya Interaction}
\author{Liang Chen}
\affiliation{Mathematics and Physics Department, North China Electric Power University,Beijing, 102206, China}
\author{Bo Hu}
\affiliation{Mathematics and Physics Department, North China Electric Power University,Beijing, 102206, China}
\author{Rong-Sheng Han}
\email[Corresponding Email: ]{hrs@ncepu.edu.cn}
\affiliation{Mathematics and Physics Department, North China Electric Power University,Beijing, 102206, China}
\date{\today}

\begin{abstract}
In this work, we use the numerical renormalization group (NRG) theory to study the thermodynamics of the two-impurity Anderson model. Two different methods are used to estimate the effect of the Dzyaloshiskii-Moriya (DM) interaction on the variation of the Kondo temperature. When the Ruderman-Kittel-Kasuya-Yosida (RKKY) interaction is vanishing, the two different estimations give different tendency. If we use the peak of the specific heat to identify the variation of the Kondo temperature versus the Dzyaloshiskii-Moriya interaction, we get an almost linear function. However, if we use the low temperature universal curve of the impurity entropy, we get a quadratic function. These results indicate that the previous debates about the influence of the spin-orbit coupling on the Kondo temperature may come from the different definitions of the Kondo temperature. When the RKKY interaction is ferromagnetic, there are two stages of the Kondo screening. Both the two estimations demonstrate that the second stage of the Kondo temperature is exponentially dependent on the DM interaction. There results are dramatically different from those calculated via perturbation theory. 

\end{abstract}


\maketitle


\section{Introduction} \label{section1} 
Magnetic impurities in a Fermi gas can give rise to the Kondo effect \cite{Hewsonbook}, a series of universal behaviors of the system induced by the quantum correlation between the itinerant electrons and the quantum impurities, below the characteristic Kondo temperature $T_K$. Magnetic doping in spin-orbit coupled systems \cite{RolandWinkler2003} has practical applications and dramatic importance, e.g., open the surface band gap \cite{Chen659,Wray07322011,PhysRevLett.106.206805,PhysRevLett.108.117601} of topological insulators \cite{RevModPhys.82.3045,RevModPhys.83.1057}, generate the quantum anomalous Hall effect in topological insulator thin films \cite{Chang167}, control the transport properties of spin-orbit coupled system \cite{CULCER2012860}, etc. However, the influence of spin-orbit coupling on the Kondo effect is far from unclear up to now. 

For the single-impurity problem, the relationship between the Kondo temperature and the spin-orbit coupling has attracted much attention. Malecki pointed out that the Kondo temperature of Kondo model  is robust against the spin-orbit coupling in two dimensional electron gas \cite{Malecki2007}. Zarea {\it et al.} reported that the Kondo temperature is enhanced by the DM interaction between the impurity spin and the two dimensional electron gas with Rashba spin-orbit coupling \cite{PhysRevLett.108.046601}. The enhancement is further confirmed by perturbative renormalization group study on the one-dimensional electron wire with Rashba and Dresselhaus spin-orbit coupling \cite{PhysRevB.94.125115}. However, both NRG calculation \cite{PhysRevB.84.193411} and quantum Monte Carlo simulation \cite{Kondo_LiangChen2016} show that the Kondo temperature is almost a linear function of the Rashba spin-orbit coupling. 

For the two-impurity model\cite{PhysRevLett.58.843,PhysRevLett.61.125,PhysRevB.40.324,PhysRevB.35.4901,PhysRevB.40.4780,PhysRevLett.72.916,PhysRevLett.68.1046,PhysRevB.52.9528,RevModPhys.80.395} in two-dimensional electron gas with Rashba spin-orbit coupling, Ivanov calculated the Kondo peaks in the even and odd channels by perturbation theory, and pointed out that the Kondo temperature has an almost-linear dependence on the spin-orbit coupling \cite{PhysRevB.86.155429}. However, as far as we know, it  is still a vacancy about the non-perturbative investigation on the Kondo temperature of two-impurity model in spin-orbit coupled systems. In the previous works, the indirect RKKY interactions between magnetic impurities in spin-orbit coupled systems have been studied broadly \cite{PhysRevLett.106.097201,PhysRevLett.106.136802,PhysRevB.81.233405,PhysRevB.89.115101}, while, the Kondo effect has not been valued. It is demonstrated that the RKKY interaction has three different terms, the isotropic Heisenberg term, the anisotropic Ising term, and the DM term\cite{PhysRevLett.106.097201}. The problem of a two-impurity model with only the isotropic Heisenberg interaction has been investigated extensively in the past two decades for the potential non-Fermi liquid behaviors in this system \cite{PhysRevLett.58.843,PhysRevLett.61.125,PhysRevB.40.324,PhysRevB.35.4901,PhysRevB.40.4780,PhysRevLett.72.916,PhysRevLett.68.1046,PhysRevB.52.9528,RevModPhys.80.395}. Garst {\it et al.} studied the model with only the Ising interaction between the two impurities, they found that there is a Kosterlitz-Thouless quantum phase transition between the singlet and the (pseudospin) doublet states \cite{PhysRevB.69.214413}. In this paper, we consider the influence of the DM interaction on the thermodynamics and the Kondo temperature of the magnetic impurities. 

The rest of the paper is organized as follows. We present the model Hamiltonian and a benchmark of the NRG calculation in Sec. \ref{section2}. In Sec. \ref{section3}, two characteristic situations are studied, the distance between the two impurities $R$ equals to $\infty$ and $\pi/2$ over the Fermi wave vector $k_F$, respectively. When $R=\infty$, the two impurities are decoupled, there is only one characteristic temperature, $T_K$, below which the two impurities are screened by the itinerant electrons. We assume that there is an additional DM interactions between the two impurities, calculate the impurity entropy and the specific heat of this model, and study the effect of the DM interaction on the variation of the Kondo temperature. When $R=\pi/2k_F$, the two impurities are ferromagnetic coupled at intermediate temperatures. There are two stages of the Kondo screening, the itinerant electrons screen the impurities from the ferromagnetic spin-1 state to the spin-1/2 state and further to the singlet state. We study the effect of the DM interaction on the characteristic temperature of the second state of the Kondo screening, $T^*$.  Finally, a discussion is given in Sec. \ref{section4}. 


\section{Model Hamiltonian}\label{section2}

In order to capture the influence of the DM interaction on the Kondo temperature, we consider the following model, 
\begin{align}
&H=\sum_{\bm{k}}\varepsilon_{\bm{k}}c_{\bm{k}}^{\dagger}c_{\bm{k}}+\varepsilon_d\sum_{j=1}^{2}\sum_{s=\uparrow,\downarrow}n_{j,s}+U\sum_{j=1}^{2}n_{j,\uparrow}n_{j,\downarrow}
\nonumber \\
&+V\left(\sum_{\bm{k},j=1,2}e^{i\bm{k}\cdot\bm{R}_j}c_{\bm{k}}^{\dagger}d_{j}+H.c.\right)+\lambda
\left(S_1^{x}S_2^{z}-S_1^{z}S_2^x\right), \label{eq1}
\end{align}
where $\epsilon_{\bm{k}}$ is the conduction band energy of state with wave-vector $\bm{k}$,  $c_{\bm{k}}^{\dagger}$ and $c_{\bm{k}}$ are the creation and the annihilation operators of the itinerant electrons. Here we consider a structureless, half-filled, noninteracting conduction band, measured from the Fermi level, ranging from $-D$ to $D$ ($D$ is the effective band width of the conduction band), such that the first term can be rewritten as, $\int_{-D}^{D}d\varepsilon\varepsilon{c^{\dagger}(\varepsilon)c(\varepsilon)}$. The index $j$ ($j=1,2$) denotes the two impurities, located at $\bm{R}_1=-\bm{R}_2=\bm{R}/2$, $\bm{R}$ is the relative distance of the two impurities. $\epsilon_d$ is the impurity energy level, $n_{j,s}=d_{j,s}^{\dagger}d_{j,s}$ is the occupation number of the $j$-th impurity state with spin $s$ ($s=\uparrow,\downarrow$). $d_{j,s}^{\dagger}$ and $d_{j,s}$ are the creation and the annihilation operators of the spin-$s$ impurity states at the $j$th location. $U$ is the Coulomb repulsion between the electrons occupying the impurity states. $V$ is the hybridization strength between the impurity states and the conduction band states.  $H.c.$ means Hermitian conjugate.  $S_{j}^{\alpha}=\frac{1}{2}d_{j,s_1}^{\dagger}\sigma^{\alpha}_{s_1,s_2}d_{j,s_2}$, $\alpha=x,y,z$, $\sigma^{x,y,z}$ are the three spin Pauli matrices. $\lambda$ is the DM interaction strength between the two local moments.

Following the standard procedure in NRG calculations \cite{RevModPhys.80.395,PhysRevB.72.104432}, we define the annihilation operators of the even and odd channels of the conduction band, 
\begin{align}
	f_{S,s}(\varepsilon)=\sum_{\bm{k}}\cos\left(\frac{\bm{k}\cdot\bm{R}}{2}\right)\frac{\delta(\varepsilon-\varepsilon_{\bm{k}})}{\sqrt{\rho(\varepsilon)\gamma_{+}(\varepsilon)}}c_{\bm{k},s}, \label{eq2}\\
	f_{A,s}(\varepsilon)=-i\sum_{\bm{k}}\sin\left(\frac{\bm{k}\cdot\bm{R}}{2}\right)\frac{\delta(\varepsilon-\varepsilon_{\bm{k}})}{\sqrt{\rho(\varepsilon)\gamma_{-}(\varepsilon)}}c_{\bm{k},s}, \label{eq3}
\end{align}
where $s=\uparrow,\downarrow$ is the spin index, $S$ and $A$ refer to the even and odd channels respectively, $\rho(\varepsilon)$ is the density of states, $\gamma_{\pm}(\varepsilon)=1\pm\frac{\sin(k_{\varepsilon}R)}{k_{\varepsilon}R}$ in three-dimensional systems, $k_{\varepsilon}$ is the wave-vector of the conduction band electrons with energy $\varepsilon$. These fermionic operators, $f_{S,s}(\varepsilon)$, $f_{A,s}(\varepsilon)$, $f_{S,s}^{\dagger}(\varepsilon)$, and $f_{A,s}^{\dagger}(\varepsilon)$, satisfy the standard anti-commutation relationships, $\{f_{I,s}(\varepsilon),f^{\dagger}_{I',s'}(\varepsilon')\}=\delta_{II'}\delta_{ss'}\delta(\varepsilon-\varepsilon')$, $I$ and $I'$ are the channel indexes. Using Eqs. (\ref{eq2})-(\ref{eq3}), we can rewrite the Hamiltonian in the energy representation, 
\begin{align}
	&H=\sum_{I=S,A}\int_{-D}^{D}d\varepsilon{\varepsilon}f_{I}^{\dagger}(\varepsilon)f_{I}(\varepsilon) \nonumber  \\
	&+\varepsilon_d\sum_{I=S,A}n_{I}+U\sum_{j=1,2}n_{j,\uparrow}n_{j,\downarrow}+\lambda\left({S_1^xS_2^z-S_1^zS_2^x}\right) \nonumber \\
	&+\sqrt{\frac{2\Gamma}{\pi}}\left(\sum_{I=S,A}\int_{-D}^{D}d\varepsilon\sqrt{\gamma_{I}(\varepsilon)}f_{I}^{\dagger}(\varepsilon)d_{I}+H.c. \right), \label{eq4}
\end{align}
where $d_{S}=(d_2+d_1)/\sqrt{2}$, $d_{A}=(d_2-d_1)/\sqrt{2}$, $n_{I}=d^{\dagger}_Id_I$, $\Gamma=\pi\rho{V}^2$ represents the hybridization strength among the impurities and the itinerant electrons. This expression (\ref{eq4}) is suitable for the NRG calculation. Before the detailed calculations, we demonstrate that both the inversion symmetry, $\bm{k}\leftrightarrow-\bm{k}$, $\bm{R}_1\leftrightarrow\bm{R}_2$, and the $SU(2)$ spin rotation symmetry are broken by the DM interaction between the two impurities. Only the $U(1)$ charge symmetry is preserved for the non-vanishing DM interaction, which makes the NRG calculation very expensive. 

\begin{figure}[tb]
	\includegraphics[width=\columnwidth]{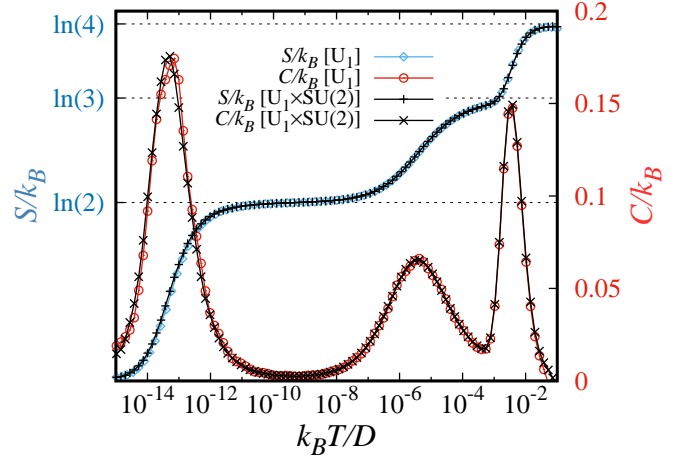} 
	\caption{(color online) Thermodynamics vs. temperature for the two-impurity Anderson model ($\lambda=0$). The blue line with "$\diamond$" and the red line with "$\circ$" refer to the impurity entropy and the impurity specific heat, respectively. They are both calculated using the $U(1)$ charge symmetry. The black lines with "$+$" and "$\times$" represent the impurity entropy and the impurity specific heat, respectively. They are calculated using the $U(1)$ charge and ${SU(2)}$ spin rotation symmetries. Parameters used in calculations are: $\Gamma=5\pi/4D$, $\varepsilon_d=-50D$, $U=100D$,  $R=\pi/2k_{F}$. The Hamiltonian is discretized with $\Lambda=5$, and the results are averaged over four $z$'s, $z=0.25$, $0.5$, $0.75$ and $1.0$.}%
	\label{fig1}%
\end{figure}

\begin{figure*}[bth]
	\includegraphics[width=0.8\textwidth]{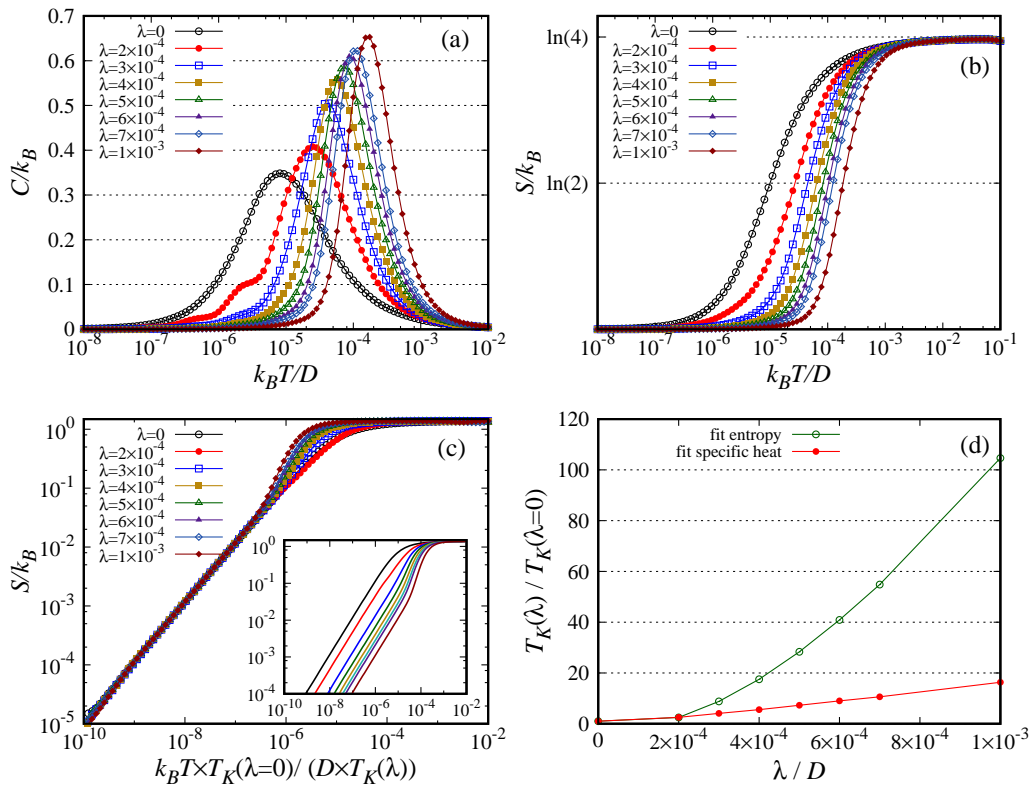}
	\caption{(color online) Thermodynamics vs. temperature for the two-impurity Anderson model for different DM interactions, $R=\infty$. (a) The impurity specific heat, $C/k_B$, vs. temperature. (b) The impurity entropy, $S/k_B$, vs. temperature. (c) The impurity entropy vs. rescaled temperature, the insert shows the unscaled impurity entropy vs. temperature in log-log coordinate. (d) The variation of Kondo temperature vs. DM interaction fitted using two different methods. The green line with "$\circ$" is fitted using the low temperature universal curve of the impurity entropy, the red line with "$\bullet$" is fitted using the peak of the impurity specific heat. Parameters used in calculation are, $\Gamma=5\pi/4D$, $\varepsilon_d=-50D$, $U=100D$. The Hamiltonian is diagonalized with $\Lambda=5$, and the results are averaged over four $z$'s, $z=0.25$, $0.5$, $0.75$ and $1.0$. }%
	\label{fig2}%
\end{figure*}

\begin{figure*}[bth]
	\includegraphics[width=0.8\textwidth]{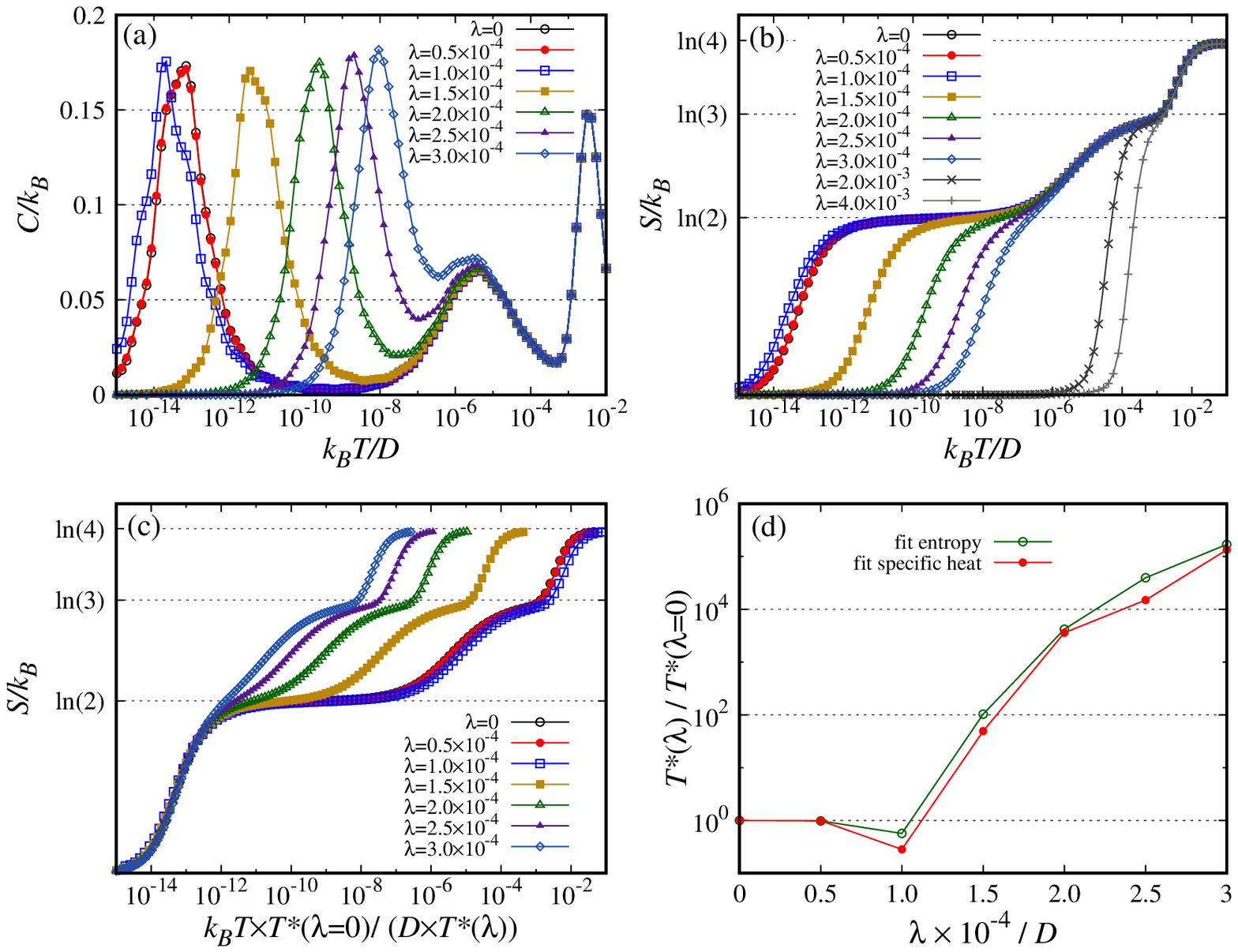}
	\caption{(color online) Thermodynamics vs. temperature for the two-impurity Anderson model for different DM interactions, $R=\pi/k_F$. (a) The impurity specific heat vs. temperature. (b) The impurity entropy vs. temperature. (c) The impurity entropy vs. rescaled temperature. (d) The variation of Kondo temperature vs. DM interaction calculated using two different methods.  The green line with "$\circ$" is fitted using the low temperature universal curve of the impurity entropy, the red line with "$\bullet$" is fitted using the peak of the impurity specific heat. Parameters used in calculation are, $\Gamma=5\pi/4D$, $\varepsilon_d=-50D$, $U=100D$. The Hamiltonian is diagonalized with $\Lambda=5$, and the results are averaged over four $z$'s, $z=0.25$, $0.5$, $0.75$ and $1.0$. }%
	\label{fig3}%
\end{figure*}

The discretization has been performed using the scheme from Ref. \onlinecite{PhysRevB.72.104432}. NRG parameters are set to be $\Lambda=5$, twist-averaging over $N_z=4$ discretization grids, $z=0.25$, $0.5$, $0.75$, 1. In each NRG calculation, $4000$ states are kept. 

Here we consider a special case for benchmark, in the limit $\lambda\rightarrow0$, the DM interaction is vanishing, the model (\ref{eq4}) tends to the traditional two impurity Anderson model, which preserves inversion symmetry, $U(1)$ charge symmetry and the $SU(2)$ spin rotation symmetry. Fig. \ref{fig1} shows the specific heat and the entropy of this special case with $R=\pi/2k_F$, calculated using different symmetries. One can find that these results are consistent to each other and those give in Ref. \onlinecite{PhysRevB.72.104432}. These results justify the validity of our calculations, and the conclusion that $4000$ states kept in each NRG iteration is sufficient.

\section{Numerical Renormalization Group Results}\label{section3}
Here we consider the case $R=\infty$. In this case, the RKKY interaction between the two impurities is vanishing. However, we assume that there may exist a non-vanishing DM interaction between the two impurities (This is experimental accessible, i.e., in spin-orbit coupled systems, the distance between the impurities is proper such that the isotropic Heisenberg like RKKY interaction is vanishing and the DM like term is non-zero \cite{PhysRevLett.106.097201}). We study the thermodynamics of the impurities via NRG method to identify the effect of DM interaction on the Kondo temperature. The black line with circles in Fig. \ref{fig2}(a) shows the specific heat of the two-impurity Anderson model (with DM interaction $\lambda=0$), where the Kondo temperature is located at $k_BT_K\approx10^{-5}D$, $k_B$ is the Boltzmann constant. One can find that it is consistent with the result given in Ref. \onlinecite{PhysRevB.70.153401}. For the increasing DM interaction from $\lambda=0$ to $\lambda/D=10^{-3}$, the peak of specific heat moves from $k_BT/D\approx10^{-5}$ to $k_BT/D\approx10^{-4}$, which demonstrate that the Kondo temperature is enhanced by the DM interaction. If we use the relative position of the peak of the specific heat to identify the variation of the Kondo temperature, i.e., the peak of the red line with dots ($\lambda/D=2\times10^{-4}$) is located at $k_BT/D=2.53\times10^{-5}$ while the black line with circles ($\lambda=0$) at $k_BT/D=8.25\times10^{-6}$, so that the Kondo temperature is enhanced for $T_K(\lambda/D=2\times10^{-4})/T_K(\lambda=0)=3.1$. The red line with dots in Fig. \ref{fig2}(d) shows the fitted results of the variation of the Kondo temperature versus the DM interaction in this method. One can find that it is almost a linear function, which is consistent with the perturbation theory results given in Ref. \onlinecite{PhysRevB.86.155429}. However, there is another method to identify the Kondo temperature via the universal curves. In the original paper \onlinecite{PhysRevB.21.1003}, the Kondo temperature is defined via the universal curve of the product of temperature and spin susceptibility versus the temperature. For the spin-orbit coupled system with DM interactions, the spin rotation symmetry is broken, spin susceptibility is generally a tensor, e.g., $\chi_{x}\ne\chi_{y}$, it is difficult to choose the proper component of the spin susceptibility as the universal curve. Here we use another low temperature universal curve, the impurity entropy \cite{Affleck1995}, to identify the variation of the Kondo temperature versus the DM interaction. Fig. \ref{fig2}(b) shows the impurity entropy versus the temperature for different DM interactions. The entropy decays from the high temperature free spin value $k_B\text{ln}(4)$ to close to zero in the region $k_BT/D\approx10^{-3}$ to $10^{-6}$. For a larger DM interaction in the parameter region $0<\lambda/D<10^{-3}$, the entropy decays to zero faster. Which demonstrates that the DM interaction between the two impurities helps the itinerant electrons to screen the magnetic impurity states, so that the Kondo temperature is enhanced by the DM interaction. As shown in the insert of Fig. \ref{fig2}(c) with log-log coordinates, in the low temperature regime ($k_BT/D\le10^{-6}$), the entropies manifest universal behaviors for different DM interactions. As given in Fig. \ref{fig2}(c), we make these curves coincide to each other in the low temperature regime by shifting the horizontal coordinates. We use the shifting ratio to define the variation of the Kondo temperature, $T_K(\lambda)/T_K(0)$, for different DM interactions. The green line with dots in Fig. \ref{fig2}(d) shows the Kondo temperature versus DM interactions via this method. One can find that the Kondo temperature is dramatically enhanced by the DM interactions. Further numerical analysis show that the ratio $T_K(\lambda)/T_K(0)$ presents quadratic dependence on the DM interactions in the parameter region $2\times10^{-4}<\lambda/D<7\times10^{-4}$. These analysis demonstrate that different definitions of Kondo temperature may lead to different conclusions about the Kondo temperature dependence on the spin-orbit coupling. 

Now we consider the other case, $R=\pi/2{k_F}$, the RKKY interaction between the two impurities is ferromagnetic. If the DM interaction is vanishing, there are three characteristic crossover of the model \cite{PhysRevB.70.153401}, from the free-impurity regime with entropy $S=k_B\text{ln}(4)$ to the FM coupled regime with entropy $S=k_B\text{ln}(3)$, then the even-channel of the conduction band screen the total spin from $1$ to $1/2$, so that the entropy decays to $S=k_B\text{ln}(2)$, finally, 1/2-spin state is screened by the odd-channel conduction band, and $S=0$. Three peaks of the specific heat (see the black line with circles in Fig. \ref{fig3}(a) for details) correspond to these crossovers. For a small DM interaction $\lambda\le3.0\times10^{-4}$, the first and second stage ($k_BT/D>10^{-6}$) of the specific heat and the entropy are hardly effected by the DM interactions. One can find that almost all of the lines are coincide to each other in this temperature region in Fig. \ref{fig3}(a) and \ref{fig3}(b). However, the screening of the total 1/2-spin state is dramatically effected ($k_BT/D<10^{-6}$). Here we present the quantitative results. We denote the characteristic temperature of the second stage of the Kondo screening as $T^*$ ($k_BT^*/D<10^{-6}$), then as same as the previous situation, we have two methods to identify the effect of the DM interaction on $T^*$. The first is using the peak of the specific heat, and the second is using the universal curves of the low temperature entropies. These results are given in Fig. \ref{fig3}(d). We find that a very small DM interaction ($\lambda/D=1\times10^{-5}$) suppresses the Kondo temperature $T^*$. Then the Kondo temperature $T^*$ is almost exponentially enhanced by the DM interactions. {\it $\lambda$ changes for only three times from $1.0\times10^{-4}D$ to $3.0\times10^{-4}D$, whereas the Kondo temperature has been changed for 5 orders of magnitude.} For a DM interaction as large as $\lambda/D\sim10^{-3}$, we find that the second state of the Kondo screening is fully suppressed by the DM interaction, and the entropy decays from $k_B\text{ln}(3)$ to 0 directly. This demonstrates that the 1/2-spin state is totally destroyed by the DM interaction if $\lambda/D\sim10^{-3}$. We have to emphasize that the ferromagnetic RKKY interaction between the impurities is as large as \cite{PhysRevB.70.153401} $-5.3\times10^{-3}D$. The characteristic DM interaction we used (i.e., $3.0\times10^{-4}D$) is much smaller than this value, such a DM interaction is experimental feasible.

\section{Discussion} \label{section4}

In this work, we use the NRG method to study the thermodynamics and the effect of the DM interaction on the Kondo temperature of the two-impurity Anderson model. We use two methods to identify the relationship between the DM interactions and the variation of the Kondo tempereture. Two situations have been considered, the first situation is the RKKY interaction induced by the itinerant electrons is vanishing. In this case, we find that different definitions of the Kondo temperature may lead to different results. if we use the peak of the specific heat to identify the variation of the Kondo temperature, we find that the Kondo temperature is almost a linear function of the DM interaction. This is qualitatively consistent with the results from perturbation theory analysis \cite{PhysRevB.86.155429}. However, if we use the low temperature universal curves to identify the variation of the Kondo temperature, we find that the Kondo temperature is almost a quadratic function of the DM interaction. These difference indicates that the previous debates on the effect of the spin-orbit coupling on the Kondo temperature may come from the different definitions of the Kondo temperature. The second situation is the RKKY interaction is ferromagnetic. There are two stages of the Kondo screening. Both the two methods demonstrate that the second stage of the Kondo screening, $T^*$, is dramatically effected by the DM interactions. An small DM interaction, which is 1 order of magnitude smaller than the ferromagnetic RKKY interaction between the two impurities, can enhance the $T^*$ for 5 orders of magnitudes. This is dramatic different from the perturbation theory results \cite{PhysRevB.86.155429}. In the previous studies, the perturbation renormalization group analysis show that the Kondo temperature can be enhanced by the DM interaction exponentially for the single-impurity Anderson model \cite{PhysRevLett.108.046601}. However, the NRG calculation \cite{PhysRevB.84.193411} and quantum Monte Carlo simulation \cite{Kondo_LiangChen2016} show that the Kondo temperature is almost a linear function of the spin-orbit energy. Some other NRG analysis find that the Kondo temperature is exponentially enhanced by the spin-orbit coupling \cite{PhysRevB.93.075148}. However, the density of states on the Fermi surface is changed by the spin-orbit coupling in their analysis. In our study on the two-impurity Anderson model, the density of state $\rho=1/2D$ is set as a constant for different DM interactions, which demonstrates that the exponentially enhancement of the Kondo temperature is essentially induced by the DM interaction rather than the alteration of the density of states on the Fermi surface. 

\section{Acknowledgement}
We great appreciate the support from the National Natural Science Foundation of China under Grant No. 11504106 and the Special Foundation for Theoretical Physics Research Program of China (Grant No. 11447167). 



\bibliographystyle{apsrev4-1}
\bibliography{ref}

\end{document}